\documentclass[a4paper,twoside]{article}

\usepackage{multirow}
\usepackage{epsfig}
\usepackage{subfigure}
\usepackage{calc}
\usepackage{amssymb}
\usepackage{amstext}
\usepackage{amsmath}
\usepackage{amsthm}
\usepackage{multicol}
\usepackage{pslatex}
\usepackage{apalike}
\usepackage{SCITEPRESS}
\usepackage[small]{caption}
\newcommand{\newatop}[2]{\genfrac{}{}{0pt}{}{#1}{#2}}
\newtheorem{theorem}{Theorem}
\newtheorem{lemma}{Lemma}
\renewcommand{\rr}[1]{{\normalfont\textrm{#1}}}
\newcommand{\bb}[1]{{\mathbb{#1}}}

\newcommand{\xsigma}{\sigma}
\newcommand{\Xt}{X(\tau)}

\begin{document}

\title{Flux through a time--periodic gate: Monte Carlo test of a 
homogenization result}

\author{\authorname{Daniele Andreucci\sup{1}, Dario Bellaveglia\sup{1}, Emilio N.M.\ Cirillo\sup{1}, and Silvia Marconi\sup{1}}
\affiliation{\sup{1}Department of Basic and Applied Sciences for Engineering, Sapienza University, v.Scarpa 16, 00161 Rome, Italy}
\email{\{daniele.andreucci, dario.bellaveglia, silvia.marconi\}@sbai.uniroma1.it, emilio.cirillo@uniroma1.it}
}

\keywords{Random Walk, Homogenization, Monte Carlo Method, Alternating Pores, Ionic Currents, 
Cooperating Evacuees}

\abstract{We investigate via Monte Carlo numerical simulations and
theoretical  considerations the outflux of random walkers moving in an
interval bounded by an interface exhibiting channels (pores, doors)
which undergo an open/close cycle according to a periodic schedule. We
examine the onset of a limiting boundary behavior characterized by a
constant ratio between the outflux and the local density, in the
thermodynamic limit. We compare such a limit with the predictions of a
theoretical model already obtained in the literature as the
homogenization limit of a suitable diffusion problem.}

\onecolumn \maketitle \normalsize \vfill

\section{\uppercase{Introduction}}
\label{s:introduzione}

\noindent 
A bunch of individuals moves at random inside a bounded
region, say the {\em playground}. 
On the boundary of the playground there are one or more {\em doors} through 
which they can exit the playground itself. 
The time average flux of individuals exiting the playground
will depend on the local density close to the doors. 
An interesting question is the following: suppose to know the rule 
governing the opening of the doors, what is the relation 
between the local individual density close to the doors and 
the outgoing flux?

This simple situation models many interesting phenomena on 
different space and time scales. We mention two examples: 
(i) the playground is a cell, the individuals are potassium 
ions, the door is a potassium channel \cite{Hille,VanDongen:2004}, 
and the problem is that 
of computing the ionic current through the channel 
\cite{ABCM2011,ABCM2012}.
This is a very important question in biology, indeed ionic channel 
are present in almost all living beings and play a key role 
in regulating the ionic concentration inside the cells. 

(ii) The playground is a smoky room (imagine a fire in a cinema), 
the individuals are evacuees, the door is the door of the room, 
and the problem is that of computing at which rate the pedestrian 
are able to escape from the room itself \cite{Armin,CM2012b,CM2012a}.
In this case the interesting problem is that of understanding 
if the way in which the evacuees behave (for instance if they cooperate 
or not) has an influence on the outgoing flux magnitude. 

In some situations, for instance when the outgoing flux is compensated 
by an incoming one, a stationary state with 
constant (in time) outgoing flux is achieved. 
In this case the ratio between the outgoing flux and the 
density close to the doors will be, obviously, a constant, which 
can be interpreted as the rate at which the individuals close to the 
doors succeed to exit the playground.
This situation is also realized on a short time scale when 
the number of individuals in the region is large with respect to the 
number of them exiting the doors per unit of time.

A different situation is that in which no incoming flux is 
present. In this case the number of individuals 
inside the playground decreases and so does the typical 
outgoing flux. The natural question is that of understanding 
if some time averaged flux has a constant ratio with respect to 
the average local density close to the door \cite{AB2011}.
This question has been posed in \cite{AB2011} under the assumption 
that the doors open with a periodic schedule. 

The setup considered in \cite{AB2011} is very basic and, hence, their 
result is absolutely general. 
A scalar field 
is defined on a $d$--dimensional open hypercube where the field 
evolves according to the diffusion equation. 
Homogeneous Neumann boundary conditions are assumed 
on the boundary of the hypercube excepting
``small" circles lying on one of the 
$(d-1)$--hypercubic faces the boundary is made of. 
In those circles the boundary condition is time--dependent on a periodic 
schedule, more precisely the positive time axis is subdivided 
in disjoint intervals (periodic cycles) 
of equal length and any of such intervals 
is subdived into two disjoint parts. 
The boundary condition on the circles is then assumed to be 
homogeneous Dirichlet into 
the first part of each of these time intervals
and homogeneous Neumann in the second part. (More general shapes than
circles are actually considered in \cite{AB2011}.)

If the field is interpreted as the density of individuals in the playgroud, 
the boundary condition in \cite{AB2011} can be described as follows: 
the boundary is always reflecting except for the small circles which 
are reflecting only in the second part of each of the time intervals 
considered above, while the individuals are allowed to exit the 
playground through these circles in the first part of each of these 
intervals. In other words the small circles are doors of the 
playground and those doors are open only in the first part of each 
of the time intervals.

The time periodic micro--structured boundary conditions
suggest to approach the problem from the homogenization theory 
point of view \cite{BLP}. With this approach in \cite{AB2011} 
it is proven that, provided the length of the open time is suitably small
with respect to the length of the cycle, the ratio between the outgoing 
flux and the 
field on the small circles (the door) is not trivial, 
in the sense that it tends to a real number when the length 
of the periodic cycles tends to zero. This constant ratio is 
explicitely computed in \cite{AB2011} and is proven to depend 
on the way in which each time interval is subdivided into two parts, 
that is to say on the length of the open door and on that of the closed 
door time sub--intervals. 
This result is, in this context, an answer to the question 
that opened the paper, namely, to the question about the 
relation between the outgoing flux and the local density of individuals
close to the exit. 

The present paper has a two--fold aim. In the one--dimensional case 
we setup a Monte Carlo 
simulation aiming to (i) test numerically the homogenization 
limiting result (in the spirit for example of \cite{Haynes}), 
(ii) compute the ratio between the outgoing flux and 
the local density close to the exit when the length of the periodic cycles
is finite. 

This project is realized by introducing a one--dimensional  discrete
space model on which independent particles perform  symmetric random
walks. The space is a finite interval on $\bb{Z}$  with a boundary point
which is reflecting, whereas the other  periodically changes its status
from absorbing to reflecting and viceversa. We tune the parameters so
that the discrete and the continuum space  models have equivalent
behaviors. Moreover, in the thermodynamics limit, namely,  when the
number of site of the discrete space model tends to infinity,  the
homogeneization result proven in the framework of the  continuum space
model is recovered.  This is not proven rigorously, but it is
demonstrated via heuristc  arguments and Monte Carlo simulations.

The paper is organized as follows. In Section~\ref{s:homo} 
we summarize the homogenization results 
found in \cite{AB2011} in the one--dimensional case.
In Section~\ref{s:rw} the discrete space model is introduced and 
its behavior is discussed on heuristic grounds. 
This model is studied via 
Monte Carlo simulations in Section~\ref{s:discu}, where all 
the numerical results are discussed.
Section~\ref{s:conc} is finally devoted to some brief conclusions. 

\section{\uppercase{A continuum space model}}
\label{s:homo}

\noindent In this section we approach the problem via a continuum space model. 
We summarize, in the one--dimensional case, 
the results found in \cite{AB2011}. We first introduce the mathematical 
model and then discuss its physical interpretation. 
 
Pick the two reals $\tau\ge\xsigma\ge0$, the integer $m$, and 
the function
$u_0\in L^2([0,L])$. 
Set $T=(m+1)\tau$ and consider the boundary value problem
consisting in the diffusion equation 
\begin{equation}
\label{diffusione}
u_t - D u_{xx} =0\;\;\;\; \textrm{ on }(0,L)\times(0,T)
\end{equation}
with $D>0$ the {\em diffusion coefficient},
the initial condition 
\begin{equation}
\label{iniziale}
u(x,0)=u_0(x)\;\;\;\;\;\forall x\in(0,L)
\end{equation}
and the boundary conditions
\begin{equation}
\label{bordo1}
u_x(0,t)=0\;\;\forall t\in[0,T)
\end{equation}
and
\begin{equation}
 \label{bordo2}
 u(L,t)=0\;\;\forall t\in A
\;\;\textrm{ and }\;\;
u_x(L,t)=0\;\;\forall t\in C
\end{equation}
where 
\begin{displaymath}
 A=\bigcup_{k=0}^{m_{\tau}}[k\tau,k\tau+\xsigma)
\;\;\;\textrm{ and }\;\;\;
 C=\bigcup_{k=0}^{m_{\tau}}[k\tau+\xsigma,k\tau+\tau)
 \,.
\end{displaymath}

According to the discussion in Section~\ref{s:introduzione}, 
the model above can be interpreted as follows: the field $u$ 
is the density of individuals in the playground,
$m$ is the number of the door opening/closing cycles,
$\tau$ is the length of each cycle, 
${\xsigma}$ is the length of the time interval in each cycle 
during which 
the door is open, and, finally, $A$ and $C$ are, 
respectively, 
the parts of the 
global time interval $[0,T)$ when the door is open
and closed.

In \cite{AB2011}, via an homogenization approach, it has 
been proven the following convergence result in the limit $\tau\to0$ for 
the solution of the boundary value problem \eqref{diffusione}--\eqref{bordo2}
providing an answer to the question about the relation between the 
individual density $u(L,t)$ at the door and the outgoing flux 
$-Du_x(L,t)$. 

\begin{theorem}
\label{t:homo}
Assume 
\begin{equation}
\label{homo01}
\exists\lim_{\tau\to0}
 \frac{\sqrt{\xsigma}}{\tau}=:\mu\ge0
\end{equation}
and let $u^{\tau}$ be the solution of the boundary value problem 
\eqref{diffusione}--\eqref{bordo2}. Then, as $\tau\to0$, $u^{\tau}$
converges in the sense of $L^2([0,L]\times[0,T))$ to the solution 
$u$ of the problem \eqref{diffusione}, \eqref{iniziale} with 
boundary conditions 
\begin{equation}
\label{homo01-bc}
u_x(0,t)=0
\;\;\;\;\forall t\in[0,T)
\end{equation}
and
\begin{equation}
\label{homo02-bc}
u_x(L,t)=-\frac{2\mu}{\sqrt{D\pi}}u(L,t)
\;\;\;\;\forall t\in[0,T)
\,.
\end{equation}
Assume 
\begin{equation}
\label{homo02}
\lim_{\tau\to0}
 \frac{\sqrt{\xsigma}}{\tau}=\infty
 \,;
\end{equation}
then 
the solution of the boundary value problem \eqref{diffusione}--\eqref{bordo2}
converges to the solution of the problem 
\eqref{diffusione}, \eqref{iniziale} with boundary condition 
\begin{equation}
\label{homo03-bc}
u_x(0,t)=
u(L,t)=0\;\;
\forall t\in[0,T)
\,.
\end{equation}
\end{theorem}

The physical meaning of the above theorem can be summarized as follows.
If the length $\tau$ of each periodic unit (cycle) is small with respect to 
$\sqrt{\xsigma}$ (see condition \eqref{homo02}), then, in the $\tau\to0$ 
limit, the system behaves as if the door were always open, namely $u(L,t)=0$.
On the other hand, 
if $\tau$ is large with respect to 
$\sqrt{\xsigma}$ 
(see condition \eqref{homo01} with $\mu=0$), then, in the $\tau\to0$ 
limit, the system behaves as if the door were always closed, namely 
$u_x(L,t)=0$.
Finally, 
if $\tau$ is of the same 
order of magnitude of 
$\sqrt{\xsigma}$ 
(see condition \eqref{homo01} with $\mu>0$), then, in the $\tau\to0$ 
limit, the system behaves as if the door were open with the outgoing 
flux constrained to satisfy the condition 
$-Du_x(L,t)=(2\mu\sqrt{D/\pi})u(L,t)$.

\subsection{A Glimpse of the Proof of Theorem~\ref{t:homo}}
\label{ss:glimpse}

In order to explain 
the mathematical meaning of the 
convergence result stated in the theorem, 
we sketch the proof of the first part of 
Theorem~\ref{t:homo}. We refer the interested reader to 
\cite{AB2011} for more details.
First of all we note that 
for the solution $u^{\tau}$ of the boundary value problem 
\eqref{diffusione}--\eqref{bordo2} it 
is not difficult to perform classical energy estimates and to prove
compactness properties in time.
Then, possibly by extracting subsequences, we have
that a function $u$ exists such that as $\tau\to0$
\begin{displaymath}
  u^{\tau} 
  \textrm{ converges strongly in }
  L^{2}([0,L]\times[0,T))
  \textrm{ to }
  u
  \,,
\end{displaymath}
and 
\begin{displaymath}
  u_x^{\tau} 
  \textrm{ converges weakly in }
  L^{2}([0,L]\times[0,T))
  \textrm{ to }
  u_x
  \,.
\end{displaymath}
Moreover, it is easily proven that $u$ satisfies 
\eqref{diffusione}--\eqref{bordo1}
in a standard weak sense. 
It is important to remark that, via these simple 
compactness considerations, it is not possible to say anything 
about the limiting boundary 
condition satisfied at $x=L$.

In order to identify such a limiting boundary condition, we consider the 
weak formulation of problem 
\eqref{diffusione}--\eqref{bordo2}.
We choose a smooth test function 
such that
\begin{displaymath}
\psi(x,t)=0
\;\;\textrm{ for }\;\;
\left\{
\begin{array}{l}
x=0 \;\textrm{ and }\; t\in(0,T)\\
x=L \;\textrm{ and }\; t\in A\\
x\in[0,L] \;\textrm{ and }\;t=T\,.\\
\end{array}
\right.
\end{displaymath}
By multiplying \eqref{diffusione} against $\psi$ and 
by integrating by parts we get
\begin{equation}
  \label{eq:testfun_weak_i}
  -
  \int_0^T
  \int_0^L
  u^{\tau}
  \psi_t
  +
  \int_0^T
  \int_0^L 
  D u^{\tau}_x
  \psi_x
  =
  \int_0^L
  u_0 
  \psi(x,0)
  \,.
\end{equation} 
Next we use the equation above with $\psi=\varphi w$, where 
$\varphi\in C^{\infty}([0,L]\times [0,T])$ is such that  
\begin{displaymath}
\varphi(x,t)=0
\;\;\textrm{ for }\;\;
\left\{
\begin{array}{l}
x=L \;\textrm{ and }\; t\in (0,T)\\
x\in[0,L] \;\textrm{ and }\;t=T\\
\end{array}
\right.
\end{displaymath}
and $w$ is chosen as follows. 

The choice of the function $w$ is the key ingredient of the proof.
Identifying the properties that the function $w$ has to satisfy in the
setting of alternating pores is   
the main point of the paper \cite{AB2011}, but the general idea of the
definition of $w$ was introduced by \cite{Friedman:1995} in a stationary case.
We consider the interval $I_\tau=(L-\sqrt{D\tau},L)$ and 
define $w$ in $I_{\tau}\times(0,T)$ as the $\tau$--periodic solution of 
the equation 
\begin{equation}
\label{eq:testfun_cond_iii}
w_t
+
D
w_{xx}
=
0
\;\;\;\;\textrm{ on }I_{\tau}\times (0,T)
\end{equation}
with boundary conditions
\begin{equation*}
w(L,t) = 0 \;\;t \in A,\;\;\;
w_x(L,t) = 0 \;\; t\in C,
\end{equation*}
and, setting for the sake of notational simplicity $\Xt=L-\sqrt{D\tau}$,
\begin{equation*}
w(\Xt,t)=1 \;\; t \in (0,T)
\,.
\end{equation*}
Notice that we extend $w=1$ for $x\in(0,\Xt)$. 
In \cite{AB2011} it is proven that as $\tau\to0$
\begin{displaymath}
  w
  \textrm{ converges strongly to $1$ in }
  L^{2}((0,L)\times(0,T))
\end{displaymath}
and 
\begin{displaymath}
  w_x 
  \textrm{ converges weakly to $0$ in }
  L^{2}([0,L]\times[0,T))
  \,.
\end{displaymath}
Moreover, it is also proven the following highly non--trivial 
property: as $\tau\to0$
\begin{multline}
 \label{eq:testfun_cond_vii}
\int_0^{T}
w_x(\Xt,t)
D
u^{\tau}(\Xt,t)
\varphi(\Xt,t) 
\to
\\
-
\frac{2\mu}{\sqrt{D\pi}}
\int_0^{T}
D 
u(L,t)
\varphi(L,t)
\,.
\end{multline}
Recall, now, equation \eqref{eq:testfun_weak_i} and notice that 
\begin{multline*}
-\int_0^{T}
\int_{0}^L 
u^{\tau} 
\varphi 
w_t
+
\int_0^{T}
\int_{0}^L 
D u^{\tau}_x
w_x
\varphi
=
\\
-
\int_0^{T}
\int_{0}^L
D u^{\tau}_x
\varphi_x 
w
+ 
\int_0^{T}
\int_{0}^L
u^{\tau}
\varphi_t 
w
\\
+   
\int_{0}^L
u_0 (x)
w(x,0) 
\varphi(x,0) 
\end{multline*}
Since 
$w$ converges strongly to $1$, we get that 
\begin{multline}
\label{eq:testfun_weak_ii} 
-\int_0^{T}
\int_{0}^L 
u^{\tau} 
\varphi 
w_t
+
\int_0^{T}
\int_{0}^L 
D u^{\tau}_x
w_x
\varphi
\;\;
\stackrel{\tau\to 0}\longrightarrow 
\\
-
\int_0^{T}
\int_{0}^L
D u_x
\varphi_x
+ 
\int_0^{T}
\int_{0}^L
u
\varphi_t  
+   
\int_{0}^L
u_0 (x) 
\varphi(x,0) 
\,.
\end{multline}
We consider next the left hand side in \eqref{eq:testfun_weak_ii} and 
compute its $\tau\to0$ limit in a different way. First of all we note 
that 
\begin{multline*}
-\int_0^{T}
\int_{0}^L 
u^{\tau} 
\varphi 
w_t
+
\int_0^{T}
\int_{0}^L 
D u^{\tau}_x
w_x
\varphi
=
\\
-
\int_0^{T}
\int_{0}^L 
u^{\tau} 
\varphi
w_t
+ 
\int_0^{T}
\int_{0}^L
D(u^{\tau} \varphi)_x 
w_x
\\
- 
\int_0^{T}
\int_{0}^L
u^{\tau} 
w_x 
\varphi_x
\,.
\end{multline*}
On the other hand, by using $(D u^{\tau} \varphi)$ as a test function for $w$ 
in \eqref{eq:testfun_cond_iii}, and integrating by parts we obtain
\begin{multline*}
-
\int_0^{T}
\int_{\Xt}^L
(D u^{\tau} \varphi)
\frac{w_t}{D}
+
\int_0^{T}
\int_{\Xt}^L
(D u^{\tau} \varphi)_x
w_x
=
\\
-\int_0^{T}
w_x(\Xt,t)
D u^{\tau}(\Xt,t)
\varphi(\Xt,t)
\,.
\end{multline*}
Recalling, now, that $w=1$ for $x\in(0,\Xt)$, from the 
two equations above we get 
\begin{multline*}
-\int_0^{T}
\!
\int_{0}^L 
u^{\tau} 
\varphi 
w_t
+
\int_0^{T}
\!
\int_{0}^L 
D u^{\tau}_x
w_x
\varphi
=
\\
-\int_0^{T}
w_x(\Xt,t)
D u^{\tau}(\Xt,t)
\varphi(\Xt,t)
\\
-
\int_0^{T}
\!
\int_{0}^L 
u^{\tau}
w_x
\varphi_x
\,.
\end{multline*}
Recalling that 
$w_x$ converges weakly to $0$ in
$L^{2}((0,L)\times(0,T))$
as $\tau\to0$, by 
\eqref{eq:testfun_cond_vii},
the above equality yields
\begin{multline}
 \label{eq:testfun_weak_v}
-\int_0^{T}
\!
\int_{0}^L 
u^{\tau} 
\varphi 
w_t
+
\int_0^{T}
\!
\int_{0}^L 
D u^{\tau}_x
w_x
\varphi
\;\;\stackrel{\tau\to 0}\longrightarrow
\\
\frac{2\mu}{\sqrt{D\pi}}
\int_0^{T}
D
u(L,t)
\varphi(L,t)
\,.
\end{multline}
By comparing 
\eqref{eq:testfun_weak_ii} and
\eqref{eq:testfun_weak_v}
we finally get
\begin{multline*}
\int_0^{T}
\int_{0}^L
[
-D
u_x
\varphi_x 
+ 
u
\varphi_t 
]
+   
\int_{0}^L
u_0(x) 
\varphi(x,0) 
\\
=
\frac{2\mu}{\sqrt{D\pi}}
\int_0^{T}
D
u(L,t)
\varphi(L,t)
\end{multline*}
which is the weak formulation of the limiting boundary flux condition 
for $u$ on $x=L$, given by
\begin{displaymath}
D
u_x(L,t)
= 
-
\frac{2\mu}{\sqrt{D\pi}}
D
u(L,t)
\qquad 
\text{for $t\in (0,T)$.}
\end{displaymath}
The theoretical approach just sketched will be commented upon also in the
Conclusions.

\section{\uppercase{A discrete space model}}
\label{s:rw}

\noindent We now approach the problem via 
a discrete space model. In this section we first define the model and then 
discuss heuristically  
the relation between the outgoing flux 
and the individual density close to the door.  
This problem will be investigated in the following section 
via Monte Carlo simulations.

We consider $N$ one--dimensional independent random walkers on 
$\Lambda=\{\ell,2\ell,\dots,n\ell\}\subset\ell\bb{Z}$ 
and denote by $t\in s\bb{Z}_+$ the time variable. 
We assume that each random walk is symmetric, only jumps between 
neighboring sites are allowed, that $0$ is a 
reflecting boundary point, and that
at the initial time the $N$ walkers are distributed 
uniformly on the set $\Lambda$. 
Moreover, we pick the two integers $1\le \bar\sigma\le \bar\tau$, we 
partition 
the time space $s\bb{Z}_+$ in  
\begin{equation*}
  A=
  \bigcup_{i=1}^\infty
  \{s(i-1)\bar\tau,\dots,s[(i-1)\bar\tau+\bar\sigma-1]\}
\end{equation*}
and
\begin{equation*}
  C=
  \bigcup_{i=1}^\infty
  \{s[(i-1)\bar\tau+\bar\sigma],\dots,s[i\bar\tau-1]\}
  \,,
\end{equation*}
and assume that the boundary point $(n+1)\ell$ is absorbing at times 
in $A$ and reflecting at times in $C$. 

More precisely, if we let $p(x,y)$ be 
the probability that the walker at site $x$ jumps to site $y$ we have that 
\begin{equation*} 
p(\ell,\ell)=\frac{1}{2},\; 
p(x,x+\ell)=\frac{1}{2}\;\textrm{ for }\;x=\ell,\dots,(n-1)\ell,
\end{equation*} 
and
\begin{equation*} 
p(x,x-\ell)=\frac{1}{2}\;\textrm{ for }\;x=2\ell,\dots,n\ell\,;
\end{equation*} 
moreover 
\begin{displaymath}
 p(n\ell,n\ell)= 
 \left\{
 \begin{array}{ll}
 0 & \textrm{at times in } A\\
 1/2 & \textrm{at times in } C\\
 \end{array}
 \right.
\end{displaymath}
and
\begin{displaymath}
 p(n\ell,(n+1)\ell)= 
 \left\{
 \begin{array}{ll}
 1/2 & \textrm{at times in } A\\
 0 & \textrm{at times in } C\,.\\
 \end{array}
 \right.
\end{displaymath}

Note that when the walker reaches the site $(n+1)\ell$ it is 
freezed there, so that this system is a model for the proposed 
problem in the following sense: each walker is an individual, 
the room is the set $\Lambda=\{\ell,\dots,n\ell\}$, 
at the initial time there are $N$ individuals in the room, 
each walker absorbed at the site $(n+1)\ell$ is counted as 
an individual which exited the room. 
We denote by $\bb{P}[\cdot]$ and 
$\bb{E}[\cdot]$ the probability and the average along the trajectories of the 
process. 

In the framework of this model an estimator for the 
ratio between the outgoing individual flux and 
the typical number of individuals close to the door 
is given by 
\begin{equation}
\label{estimatore}
K_i=
\frac{\bb{E}[F_i]/(s\bar\tau)}{(\bb{E}[U_i]/\bar{\tau})/\ell}
\;\;\;\;\;\;\textrm{ for all }i\in\bb{Z}_+
\end{equation}
where
$F_i$ is the number of walkers that reach the boundary point $(n+1)\ell$ 
during the $i$--th cycle,
$U_i$ is the sum over the time steps in the $i$--th cycle 
of the number of walkers at the site $\ell n$.

We are interested into two main problems. 
The first question that we address is the dependence 
on time of the above ratio, in other words we wonder if this quantity 
does depend on $i$.
The second problem that we investigate is the connection between the 
predictions of this discrete time model and those provided by 
the continuous space one introduced in 
Section~\ref{s:homo}.
These two problems will be discussed in this section via 
heuristic estimates and in the next one via Monte Carlo simulations.
Both analytic and numerical computations will be performed under 
the assumptions 
\begin{equation}
\label{assumo}
\bar\tau\gg\bar\sigma
\;\;\;\textrm{ and }\;\;\;
n>2\bar\sigma
\,.
\end{equation}
The first hypothesis says that the time interval in which the 
right hand boundary point is absorbing is much smaller than
that in which it is reflecting. In other words in each cycle
the door is open in a very short time subinterval.
The second assumption says that the lenght of the space interval 
is larger than $2\bar\sigma$ and this 
will ensure that particles being absorbed by the right hand 
boundary in a given cycle do not feel the presence of the left hand endpoint
in that cycle.

\subsection{The estimator $K_i$ is a constant}
\label{s:time}
\par\noindent
Under 
the first of the two assumptions \eqref{assumo}, it is reasonable 
to guess that during any cycle the walkers in the system 
are distributed uniformly in $\Lambda$, so that at each time and 
at each site of $\Lambda$ the number of walker on that site 
is approximatively given by $\bb{E}[U_i]/\bar\tau$. 
Since $\bar\sigma$ is much smaller than $\bar\tau$, the mean number of 
walkers $\bb{E}[F_i]$ that reach the boundary point $(n+1)\ell$ during the 
cycle $i$ is proportional to $\bb{E}[U_{i-1}]/\bar\tau$ and the 
constant depends only on $\bar\sigma$, so that we have 
\begin{equation}
\label{alpha}
\bb{E}[F_i]=\frac{\alpha(\bar\sigma)}{\bar\tau}\,\bb{E}[U_{i-1}]
\,.
\end{equation}
We also note that, since $\bar\tau\gg\bar\sigma$, we have that 
\begin{displaymath}
n\frac{1}{\bar\tau}\bb{E}[U_i]
=
n\frac{1}{\bar\tau}\bb{E}[U_{i-1}]
-\bb{E}[F_i]
\end{displaymath}
By combining the two equations above we get that 
\begin{equation}
\label{estimatore02}
K_i
=
K
\equiv
\Big[\frac{1}{\alpha(\bar\sigma)}-\frac{1}{n}\Big]^{-1}
\frac{1}{\bar\tau}
\frac{\ell}{s}
\end{equation}
showing that the estimator \eqref{estimatore} 
does not depend on time, namely, it is equal to $K$ for each $i$. 

\subsection{Estimating $\alpha(\bar\sigma)$}
\label{s:alpha}
\par\noindent
As it will be discussed in the following subsection, we are 
interested in finding an estimate for 
$\alpha(\bar\sigma)$ in the limit $\bar\sigma$ large. 

First of all we give a very rough estimate of such a constant. 
As noted above, since we assumed, 
$\bar\tau\gg\bar\sigma$, it is reasonable 
to imagine that the walkers are distributed uniformly 
with density $\bb{E}[U_{i-1}]/\bar\tau$ when the $i$--th cycle 
begins (opening of the door).
Hence, since the walkers are independent, we get 
\begin{displaymath}
\bb{E}[F_i]
=
\frac{\bb{E}[U_{i-1}]}{\bar\tau}
\times 
S
\,,
\end{displaymath}
where
we denote by $S$
the sum over the particles that at time $(i-1)\bar\tau-1$ 
are less than $\bar\sigma$ sites from 
the absorbing boundary point of the probability
that each of them reaches the absorbing boundary 
in the next $\bar\sigma$ time steps.
Recalling \eqref{alpha}, we have
\begin{equation}
\label{alpha02}
\alpha(\bar\sigma)
=
S
\,.
\end{equation}

This representation allows an immediate rough estimate of the quantity 
$\alpha(\bar\sigma)$. 
If $\bar\sigma$ is large, 
at time $\bar\sigma$ each walker space distribution probability 
can be approximated by a gaussian function with variance 
$\sqrt{2 \bar\sigma}$ (Central Limit Theorem). 
Hence, the number of particles that reach in the following 
$\bar\sigma$ steps the boundary $(n+1)\ell$ is 
approximatively given by the number of walkers at the 
$\sqrt{2 \bar\sigma}$ sites counted starting from the absorbing 
boudary point divided by $2$. Hence, we find the estimate 
\begin{displaymath}
\alpha(\bar\sigma)
\approx
\frac{1}{2}\sqrt{2\bar\sigma}
=
\sqrt{\frac{\bar\sigma}{2}}
\end{displaymath}
suggesting that 
the quantity $\alpha(\bar\sigma)$ depends on $\bar\sigma$
as $\sqrt{\bar\sigma}$.

We now discuss a more precise argument.
In order to compute the right hand term in \eqref{alpha02}
we consider a particle performing a simple symmetric random walk 
on $\bb{Z}$ and denote by $\bb{Q}$ the probability along the 
trajectories of the process. Since we have assumed 
$n>2\bar\sigma$, see \eqref{assumo}, 
the probability that a particle in the 
original model starting at a position which is $y$ site far   
from the absorbing boundary point, with 
$1\le y\le\bar\sigma$,
reaches such a point in a time smaller than or equal to 
$\bar\sigma$ is equal to the 
probability that the single symmetric walker on $\bb{Z}$ 
starting at $0$ reaches the point $y$ in a time 
smaller than or equal to $\bar\sigma$.
Then, if we let $T_y$ be the first hitting time to $y\in\bb{Z}$
for the simple symmetric walker on $\bb{Z}$ started at $0$, from
\eqref{alpha02}, we have that 
\begin{multline*}
\alpha(\bar\sigma)
=
\sum_{y=1}^{\bar\sigma} \bb{Q}[T_y\le\bar\sigma]
=
\sum_{y=1}^{\bar\sigma} 
\sum_{h=y}^{\bar\sigma} 
  \bb{Q}[T_y=h]
  \\
=
\sum_{y=1}^{\bar\sigma} 
\sum_{h=y}^{\bar\sigma} 
\frac{y}{h}\bb{Q}[S_h=y]
\end{multline*}
where $S_h$ denotes the position of the walker at time $h$ and 
in the last equality we have used 
\cite[Theorem~14 in Section~3.10]{GS}.
Recalling, now, \cite[equation~(2) in Section~3.10]{GS}, we have that 
\begin{equation}
\label{alpha04}
\alpha(\bar\sigma)
=
\sum_{y=1}^{\bar\sigma} 
y
\sum_{\newatop{h=y:}{h+y\textrm{ even}}}^{\bar\sigma} 
\frac{1}{h}
\binom{h}{(h+y)/2}\frac{1}{2^h}\,.
\end{equation}

We first remark that, since $\alpha(\bar\sigma)$ is 
a double sum of positive terms, we have that 
$\alpha(\bar\sigma)$ is an increasing function of 
$\bar\sigma$. In the next theorem we state two important properties 
of $\alpha(\bar\sigma)$. 
The proof of the theorem will use the result stated in the 
following lemma.

\begin{lemma}
\label{t:stolzcesaro}
Let $f:\bb{Z}_+\to\bb{R}$ be a function such that the limit 
$\lim_{m\to\infty}f(m)$
does exist. Then, 
\begin{displaymath}
\lim_{m\to\infty}
\frac{1}{\sqrt{m}}\sum_{i=1}^m\frac{1}{\sqrt{i}}f(i)
=
2\lim_{m\to\infty}f(m)
\end{displaymath}
\end{lemma}

\medskip
\par\noindent
\textit{Proof.\/} 
First note that 
\begin{multline*}
\lim_{m\to\infty}
\frac{1}{\sqrt{m+1}-\sqrt{m}}
 \Big[
      \sum_{i=1}^{m+1}\frac{1}{\sqrt{i}}f(i)
      -
      \sum_{i=1}^m\frac{1}{\sqrt{i}}f(i)
 \Big]
 \\
=
\lim_{m\to\infty}
\frac{1}{\sqrt{m+1}-\sqrt{m}}
      \frac{f(m+1)}{\sqrt{m+1}}
=
2
\lim_{m\to\infty}
f(m)
\end{multline*}
The statement follows by the Stolz-Ces\`aro theorem.
\qed

\begin{theorem}
\label{t:alpha}
The function $\alpha:\bb{Z}_+\to\bb{R}$ satisfies
\begin{equation}
\label{alpha-stat01}
\lim_{r\to\infty}
\frac{\alpha(r)}{\sqrt{r}}
=\sqrt{\frac{2}{\pi}}
\,.
\end{equation}
\end{theorem}

\medskip
\par\noindent
\textit{Proof.\/}
We assume $r$ even; the case $r$ odd can be treated similarly.  
In order to get \eqref{alpha-stat01} we 
rewrite \eqref{alpha04} as 
\begin{equation}
\label{alpha07}
\alpha(r)
=
\alpha_\rr{e}(r)
+
\alpha_\rr{o}(r)
\end{equation}
with 
\begin{displaymath}
\alpha_\rr{e}(r)
\equiv
\sum_{k=1}^{r/2}
(2k)
\sum_{s=k}^{r/2}
\frac{1}{2s}
\binom{2s}{(2s+2k)/2}\frac{1}{2^{2s}}
\end{displaymath}
and 
\begin{multline*}
\alpha_\rr{o}(r)
\equiv
\sum_{k=1}^{r/2}
(2k-1)
\\
\times
\sum_{s=k}^{r/2}
\frac{1}{2s-1}
\binom{2s-1}{(2s+2k-2)/2}\frac{1}{2^{2s-1}}
\end{multline*}
We shall prove that 
\begin{equation}
\label{alpha08}
\lim_{r\to\infty}
\frac{\alpha_\rr{e}(r)}{\sqrt{r}}
=\sqrt{\frac{1}{2\pi}}
\,;
\quad
\lim_{r\to\infty}
\frac{\alpha_\rr{o}(r)}{\sqrt{r}}
=\sqrt{\frac{1}{2\pi}}
\end{equation}
and hence \eqref{alpha07} will imply \eqref{alpha-stat01}. 

We are then left with the proof of \eqref{alpha08}. We only prove the first of 
the two limits; the argument leading to the second one is similar.
First of all we note that
\begin{multline*}
\alpha_\rr{e}(r)
=
\sum_{k=1}^{r/2}
\sum_{s=k}^{r/2}
\frac{k}{s}
\binom{2s}{s+k}\frac{1}{2^{2s}}
\\
\phantom{mi}
=
\sum_{s=1}^{r/2}
\sum_{k=1}^{s}
\frac{k}{s}
\binom{2s}{s+k}\frac{1}{2^{2s}}
=
\sum_{s=1}^{r/2}
\sum_{h=s+1}^{2s}
\frac{h-s}{s}
\binom{2s}{h}\frac{1}{2^{2s}}
\end{multline*}
Thus, by using the properties of the binomial coefficients we get
\begin{multline*}
\!\!
\!\!
\!\!
\!
\alpha_\rr{e}(r)
=
-
\sum_{s=1}^{r/2}
\sum_{h=s+1}^{2s}
\binom{2s}{h}\frac{1}{2^{2s}}
+
\sum_{s=1}^{r/2}
\sum_{h=s+1}^{2s}
\frac{h}{s}
\binom{2s}{h}\frac{1}{2^{2s}}
\\
\phantom{m}
=
-
\sum_{s=1}^{r/2}
\sum_{h=s+1}^{2s}
\binom{2s}{h}\frac{1}{2^{2s}}
+
\sum_{s=1}^{r/2}
\sum_{h=s+1}^{2s}
\binom{2s-1}{h-1}\frac{1}{2^{2s-1}}
\end{multline*}
and, hence, 
\begin{multline*}
\alpha_\rr{e}(r)
=
-
\sum_{s=1}^{r/2}
\sum_{h=s+1}^{2s}
\binom{2s}{h}\frac{1}{2^{2s}}
\\
+
\sum_{s=1}^{r/2}
\sum_{\ell=s}^{2s-1}
\binom{2s-1}{\ell}\frac{1}{2^{2s-1}}
\end{multline*}
Now, by the Newton's binomial theorem we get 
\begin{multline}
\label{alpha11}
\alpha_\rr{e}(r)
=
\sum_{s=1}^{r/2}
\Big\{
      -\frac{1}{2}\Big[1-\frac{1}{2^{2s}}\binom{2s}{s}\Big]+\frac{1}{2}
\Big\}
\\
=
\sum_{s=1}^{r/2}
     \frac{1}{2^{2s+1}}\binom{2s}{s}
\end{multline}
which is a notable expression for $\alpha_\rr{e}$.
The Stirling's approximation finally yields
\begin{multline*}
\alpha_\rr{e}(r)
=
\sum_{s=1}^{r/2}
     \frac{1}{2^{2s+1}}
     2^{2s}\frac{1}{\sqrt{\pi}}\frac{1}{\sqrt{s}}[1+g(s)]
     \\
=
\frac{1}{2\sqrt{\pi}}
\sum_{s=1}^{r/2}
     \frac{1}{\sqrt{s}}[1+g(s)]
\end{multline*}
where $g(s)\to0$ as $s\to\infty$. Hence, 
\begin{multline}
\label{alpha12}
\lim_{r\to\infty}
\frac{\alpha_\rr{e}(r)}{\sqrt{r}}
=
\frac{1}{2\sqrt{\pi}}
\lim_{r\to\infty}
\frac{1}{\sqrt{r}}
\sum_{s=1}^{r/2}
     \frac{1}{\sqrt{s}}[1+g(s)]
     \\
=
\frac{1}{2\sqrt{2\pi}}
\lim_{t\to\infty}
\frac{1}{\sqrt{t}}
\sum_{s=1}^{t}
     \frac{1}{\sqrt{s}}[1+g(s)]
\end{multline}
The first of the two limits \eqref{alpha08} finally follows from 
\eqref{alpha12} and Lemma~\ref{t:stolzcesaro}.
\qed

Moreover, also relying upon the numerical simulations, 
we conjecture that there exists a positive integer $r_0$ such that 
\begin{equation}
\label{conj}
\frac{\alpha(r+1)}{\sqrt{r+1}}
-
\frac{\alpha(r)}{\sqrt{r}}
>0
\end{equation}
for any integer $r\ge r_0$.

\subsection{Comparison with the continuum space model}
\label{s:continuo}
\par\noindent
In order to compare the results discussed above in this section with those in 
Section~\ref{s:homo} referring to the continuous space 
model defined therein, we have to consider two limits. 
The parameter $\bar\sigma$ has to be taken large (recall, 
also, that we always assume $\bar\tau\gg\bar\sigma$,
see \eqref{assumo})
so that, 
due to the Central Limit Theorem,  
the discrete and the continuous space model have similar behaviors 
provided
the other parameters are related as 
$2Ds=\ell^2$.
With this choice of the parameters, then, 
we expect that, 
provided the ratio $\sigma/\tau^2$ is chosen properly,
the discrete space model 
will give results similar to those predicted by the 
continuous space one with finite $\tau$. 

In~\cite{AB2011}, see Theorem~\ref{t:homo}, the relation between 
the outoing flux and the density close to the pore is worked out only 
in the limit $\tau\to0$. We then have to understand how to implement such 
a limit in our discrete time model. 

We perform this analysis in the critical case $\xsigma=\mu^2\tau^2$.
In order to compare the discrete and the continuum space models we first let 
\begin{equation}
\label{lenght}
\ell=\frac{L}{n+1}
\,.
\end{equation}
As already remarked above, 
from the Central Limit Theorem, it follows 
that the two models give the same long time predictions if 
$2Ds=\ell^2$; hence, the time unit is set to 
\begin{equation}
\label{time}
s=\frac{\ell^2}{2D}=\frac{L^2}{2 D (n+1)^2}
\,.
\end{equation}
We then consider the random walk model introduced above 
by choosing $\bar\sigma$ and $\bar\tau$ such that 
the equality 
$\bar\sigma s = (\mu \bar\tau s )^2$ 
is satisfied as closely as possible (note that $\bar\tau$ and $\bar\sigma$ 
are integers). 
This can be done as follows: we fix $L$, $n$, $\mu$, and $\bar\sigma$ and 
we then consider 
\begin{equation}
\label{ntau}
\bar\tau
=
\bigg\lfloor
\frac{1}{\mu}
\sqrt{\frac{\bar\sigma}{s}}
\bigg\rfloor
=
\frac{1}{\mu}
\frac{n+1}{L}
\sqrt{2D\bar\sigma}
-\delta
\end{equation}
where $\lfloor\cdot\rfloor$ denotes the integer part of a real number and 
$\delta\in[0,1]$. 
With the above choice of the parameters, the behavior of the 
random walk model has to be compared with that of the continuum space 
model in Section~\ref{s:homo} with period
\begin{equation}
\label{tau}
\tau
=
s\bar\tau
=
\frac{1}{\mu}
\frac{L\sqrt{\bar\sigma}}{(n+1)\sqrt{2D}}
-
\frac{L^2}{2D(n+1)^2}
\delta
\,.
\end{equation}

The equation \eqref{tau} is very important in our computation, since it 
suggests that the homogenization limit $\tau\to0$ studied in the 
continuum model should be captured by the discrete space model 
via the thermodynamics limit $n\to\infty$. 
We then expect that the estimator $K$ has to converge to the constant 
$2\mu\sqrt{D}/\sqrt{\pi}$ in this limit. 

This seems to be the case if we use the 
heuristic estimate of 
the constant $K$
obtained above. Indeed, by \eqref{estimatore02} and \eqref{tau}, 
we have that
\begin{multline}
\label{estimatore03}
K
=
\bigg[\frac{1}{\alpha(\bar\sigma)}-\frac{1}{n}\bigg]^{-1}
\sqrt{\frac{2D}{\bar\sigma}}\mu
\\
\times
\bigg[
      1+\frac{\delta\mu L}{\sqrt{2D\bar\sigma}}\frac{1}{n+1}
       +o\bigg(\frac{1}{n+1}\bigg)
\bigg]
\end{multline}
for the ratio between the outgoing flux and the local density close 
to the door, where $o(1/(n+1))$ is a function tending to zero faster 
than $1/(n+1)$ in the limit $n\to\infty$. 
In the next section we shall obtain such an estimate via a 
Monte Carlo computation, but here, by using (\ref{alpha-stat01}), 
we get that
\begin{displaymath}
K
\stackrel{n\to\infty}{\longrightarrow}
\frac{\alpha(\bar\sigma)}{\sqrt{\bar\sigma}}
\sqrt{2D}\mu
\stackrel{\bar\sigma\to\infty}{\longrightarrow}
\sqrt{\frac{2}{\pi}}
\sqrt{2D}\mu
=2\mu\sqrt{\frac{D}{\pi}}
\end{displaymath}
which is the desired limit.

\section{Monte Carlo results}
\label{s:discu}
\par\noindent
In this section we describe the Monte Carlo computation of the 
constant \eqref{estimatore}. This measure is quite difficult since 
in this problem the stationary state is trivial, in the sense 
that, since there is an outgoing flux through the boudary point $(n+1)\ell$ 
and no ingoing flux is present, all the particles will eventually exit 
the system itself. 

\begin{table*}[t]
\caption{The parameter $\tau$, computed via \eqref{tau}, for the specified 
         values of $\bar\sigma$ and $n$.}
\label{t:tau} 
\centering
\begin{tabular}{cc|c|c|c|c|c|c|c|c|c|}
\cline{3-11}
& & \multicolumn{9}{ c| }{$n$} \\ 
\cline{3-11}
& & 200 & 400 & 600 & 800 & 1000 & 1500 & 3000 & 5000 & 10000\\ 
\cline{1-11}
\multicolumn{1}{ |c| }{\multirow{3}{*}{$\bar\sigma$} } &
\multicolumn{1}{ |c| }{30} & 0.0865 & 0.0431 & 0.0287 & 0.0215  
                           & 0.0172 & 0.0115 & 0.0057 & 0.0034 & 0.0017 \\
\cline{2-11}
\multicolumn{1}{ |c }{} &
\multicolumn{1}{ |c| }{100} & 0.1579 & 0.0787 & 0.0524 & 0.0393  
                            & 0.0314 & 0.0210 & 0.0105 & 0.0063 & 0.0031 \\
\cline{2-11}
\multicolumn{1}{ |c }{} &
\multicolumn{1}{ |c| }{200} & 0.2233 & 0.1114 & 0.0742 & 0.0556   
                            & 0.0445 & 0.0296 & 0.0148 & 0.0089 & 0.0044 \\
\cline{1-11}
\end{tabular}
\end{table*}

\begin{table*}[t]
\caption{Measured constant $K$ for the specified values of $\bar\sigma$ 
         and $n$.}
\label{t:costante} 
\centering
\begin{tabular}{cc|c|c|c|c|c|c|c|c|c|}
\cline{3-11}
& & \multicolumn{9}{ c| }{$n$} \\ 
\cline{3-11}
& & 200 & 400 & 600 & 800 & 1000 & 1500 & 3000 & 5000 & 10000\\ 
\cline{1-11}
\multicolumn{1}{ |c| }{\multirow{3}{*}{$\bar\sigma$} } &
\multicolumn{1}{ |c| }{30} & 0.8660 & 0.8140 & 0.7916 & 0.7794 
                           & 0.7723 & 0.7624 & 0.7476 & 0.7371 & 0.7351 \\
\cline{2-11}
\multicolumn{1}{ |c }{} &
\multicolumn{1}{ |c| }{100} & 1.0059 & 0.9099 & 0.8772 & 0.8559 
                            & 0.8430 & 0.8245 & 0.8017 & 0.7906 & 0.7772  \\
\cline{2-11}
\multicolumn{1}{ |c }{} &
\multicolumn{1}{ |c| }{200} & 1.1135 & 0.9738 & 0.9269 & 0.8994 
                            & 0.8852 & 0.8564 & 0.8280 & 0.8155 & 0.7944 \\
\cline{1-11}
\end{tabular}
\end{table*}

Our problem can be rephrased as follows: both the outgoing 
flux and the local density at the door are two ``globally decreasing''
random variables, but their mutual ratio is constant in average. 
We then have to set up a procedure to capture this constant ratio. 

For the time length of the open state,
we shall consider the following values 
\begin{displaymath}
\bar{\sigma}
=
30, 50, 70, 100, 120, 150, 200
\,.
\end{displaymath}
For each of them, in order to 
perform the limit $\tau\to0$, we shall consider 
\begin{displaymath}
n
=
200, 400, 600, 800, 1000, 1500, 3000, 5000, 10000
\end{displaymath}
for the number of sites of the lattice $\Lambda$.

For each choice of the two parameters $\bar{\sigma}$ and $n$ we shall 
run the process and compute at each cycle $i$ the quantity 
\begin{displaymath}
k_i
=
\frac{F_i/(\bar\tau)}{U_i/(\bar\tau)}
\end{displaymath}
where, we recall, $\bar\tau$ is defined in \eqref{ntau} and 
$F_i$ and $U_i$ have been defined below \eqref{estimatore}.

\begin{figure}[!h]
  \vspace{-1.7cm}
  \centering
   {\epsfig{file = 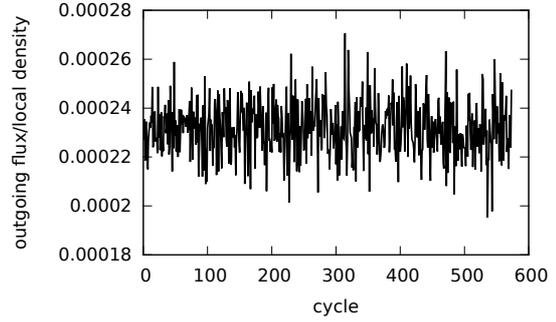, width = 10.5cm}}
  \caption{The quantity $k_i$ is plotted vs.\ the cycle number $i$ in the 
           case $\bar\sigma=30$ and $n=5000$.}
  \label{f:kiuno}
  \vspace{-0.1cm}
\end{figure}

\begin{figure}[!h]
  \vspace{-1.7cm}
  \centering
   {\epsfig{file = 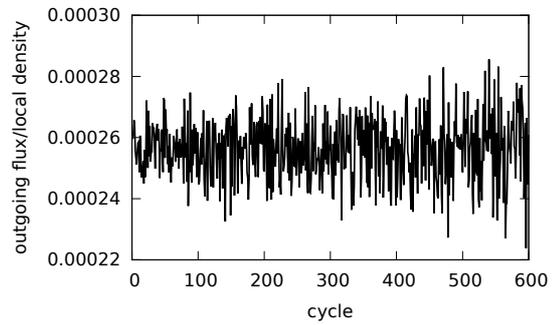, width = 10.5cm}}
  \caption{The quantity $k_i$ is plotted vs.\ the cycle number $i$ in the 
           case $\bar\sigma=200$ and $n=5000$.}
  \label{f:kidue}
  \vspace{-0.1cm}
\end{figure}

The quantity $k_i$ is a random variable 
fluctuating with $i$, but, as it is illustrated in the 
Figures~\ref{f:kiuno} and \ref{f:kidue}, it performs random 
oscillations around a constant reference value. 
We shall measure this reference value by computing 
the time average of the quantity $k_i$. We shall average $k_i$ by 
neglecting the very last cycles which are characterized by large 
oscillations due to the smallness of the number of residual particles in 
the system. 

The product of the reference value for the random variable $k_i$ 
and the quantity $\ell/s$, see the equations \eqref{estimatore}, 
\eqref{lenght} and \eqref{time},
will be taken as 
an estimate for $K$.  In other words the output of our computation will be the quantity 
\begin{equation}
\label{costK}
K=
\frac{\ell}{s}\times (k_i\textrm{ time average}) 
\,.
\end{equation}

\begin{figure}[!h]
  \vspace{-1.7cm}
  \centering
   {\epsfig{file = 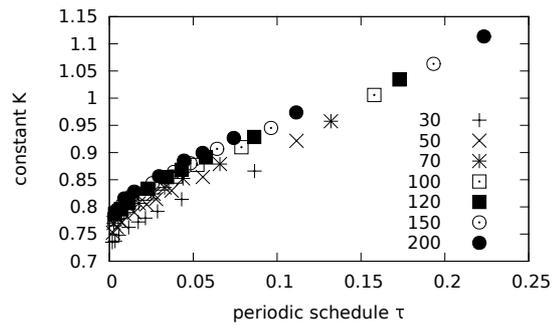, width = 10.5cm}}
  \caption{The Monte Carlo estimate of the constant $K$ measured as in  
           \eqref{costK} vs.\ the periodic time schedule $\tau$. Each 
            series of data refers to the $\bar\sigma$ value reported on
            the right bottom part of the figure.}
  \label{f:costante}
  \vspace{-0.1cm}
\end{figure}

\begin{figure}[!h]
  \vspace{-1.7cm}
  \centering
   {\epsfig{file = 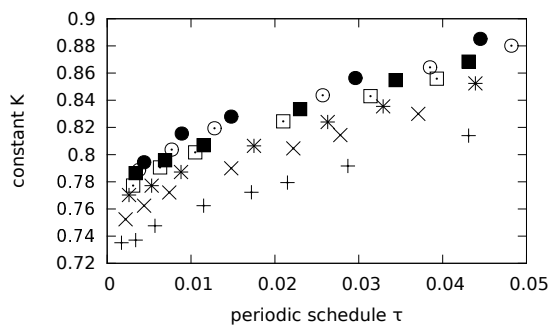, width = 10.5cm}}
  \caption{The same data as in figure~\ref{f:costante} zoomed in the 
           interval $[0,0.05]$.}
  \label{f:costante2}
  \vspace{-0.1cm}
\end{figure}

We perform the computation described 
above with $D=1$, $L=\pi$, $\mu=1/\sqrt{2}$; 
with this choice the continuum space model prediction for the 
ratio is $2\mu\sqrt{D}/\sqrt{\pi}=0.798$.

Our numerical results are illustrated in Figures~\ref{f:costante}
and \ref{f:costante2}.
We note that by increasing $\bar\sigma$ the numerical series tend to collapse 
to one limiting behavior. This is in agreement with what we proved in 
Section~\ref{s:alpha}. Moreover, provided $\bar\sigma$ is large 
enough, for $\tau\to0$ the 
measured constant tends to the theoretical value $0.798$.
For $\bar\sigma=30,100,200$ 
we have also reported in Tables~\ref{t:tau} and \ref{t:costante} the data 
plotted in Figure~\ref{f:costante}.

We can finally state that the Monte Carlo measure of the constant $K$
is in very good agreement with the theoretical predictions discussed above. 

We also note that, both the continuum space study outlined in 
Section~\ref{s:homo} and the heuristic discussion of its discrete space 
counterpart given in Section~\ref{s:rw} were just able to predict the 
value of the constant $K$ in the limit $\tau\to0$. 
No information was given on its behavior at finite $\tau$. 

The Monte Carlo computations, on the other hand, suggest that $K$
increases  with the periodic schedule $\tau$. We cannot give, at this
stage of our  reasearch, a physical interpretation of this result. This
is for sure a  very interesting point in the framework of this problem,
indeed, it is  connected with the efficiency of the evacuation
phenomenon in connection  with the periodicity of the open/close door
cycles.

\section{Conclusions}
\label{s:conc}

We have studied via Monte Carlo simulations the outgoing flux through a 
``door'' periodically alternating between open and closed states.
We have shown that the discrete space random walk model exhibits the onset
of the same limiting behavior as the continuum space model sketched
in Section~\ref{s:homo}. The homogenization limit of the continuum space 
model corresponds to the thermodynamics limit in the discrete space one.

The first one of the goals stated in the Introduction, that is the numerical
test of the homogenization result, has been in our opinion achieved (see the
Figures and the comments in Section~\ref{s:discu}). We remark that we raised
some problems in the theory of random walk which, albeit not tackled in this
paper, seem to deserve a theoretical investigation (see Section~\ref{s:time}).

As to our second goal of investigating the problem for finite $\tau$, we
have found clear evidence of a monotonic behavior in $\tau$ of the
estimator $K$, which we deem believable in view of the just commented
coherence shown by the Monte Carlo method with the theoretical
Theorem~\ref{t:homo}.

In this connection we must remark that even  from the short account of
the main steps in the proof of Theorem~\ref{t:homo}, given in
Section~\ref{ss:glimpse}, it is quite clear that the monotonic behavior
identified by the Monte Carlo approach is not easily amenable to
investigation, or even discovery, by means of that theoretical approach.

As remarked in the previous Section, we do not presently provide a full
insight in the origin and meaning of this behavior, which however is
connected with our conjecture \eqref{conj}, and with the efficiency of
the evacuation phenomenon as a function of $\tau$. It is important to
recall, finally, that at least in biological applications the efficiency
of this mechanism is not the only concern. For example the alternating schedule
of ion channels has been connected to the selection of a preferred ion species
\cite{VanDongen:2004}. Thus in general we expect $\tau$ to satisfy
several different constraints coming from different features of the biological
system.

%\begin{acknowledgements}
%\end{acknowledgements}
\vfill
\bibliographystyle{apalike}
{\small
\bibliography{abcm_simul}}
\end{document}